\documentclass[reprint,amsmath,amssymb,aps]{revtex4-1}
\usepackage{graphicx}
\usepackage{dcolumn}
\usepackage{bm}
\usepackage{hyperref}
\usepackage[mathlines]{lineno}

\begin{document}

\preprint{APS/Quant_Metro}

\title{Achieving Heisenberg-limited metrology with spin cat states via interaction-based readout}

\author{Jiahao Huang$^{1}$}
\author{Min Zhuang$^{1,2}$}
\author{Bo Lu$^{1}$}
\author{Yongguan Ke$^{1,2}$}
\author{Chaohong Lee$^{1,2,3}$}
\altaffiliation{Corresponding author.\\ Email: lichaoh2@mail.sysu.edu.cn, chleecn@gmail.com}

\affiliation{$^{1}$Laboratory of Quantum Engineering and Quantum Metrology, School of Physics and Astronomy, Sun Yat-Sen University (Zhuhai Campus), Zhuhai 519082, China}

\affiliation{$^{2}$State Key Laboratory of Optoelectronic Materials and Technologies, Sun Yat-Sen University (Guangzhou Campus), Guangzhou 510275, China}

\affiliation{$^{3}$Synergetic Innovation Center for Quantum Effects and Applications, Hunan Normal University, Changsha 410081, China}
\date{\today}

\begin{abstract}
Spin cat states are promising candidates for quantum-enhanced measurement.
Here, we analytically show that the ultimate measurement precision of spin cat states approaches the Heisenberg limit, where the uncertainty is inversely proportional to the total particle number.
In order to fully exploit their metrological ability,  we propose to use the interaction-based readout for implementing phase estimation.
It is demonstrated that the interaction-based readout enables spin cat states to saturate their ultimate precision bounds.
The interaction-based readout comprises a one-axis twisting, two $\frac{\pi}{2}$ pulses, and a population measurement, which can be realized via current experimental techniques.
Compared with the twisting echo scheme on spin squeezed states, our scheme with spin cat states is more robust against detection noise.
Our scheme may pave an experimentally feasible way to achieve Heisenberg-limited metrology with non-Gaussian entangled states.

\end{abstract}


\maketitle

\section{Introduction}\label{Sec1}
Quantum metrology aims to enhance the measurement precision and develop optimal schemes for estimating an unknown parameter by means of quantum strategies~\cite{Giovannetti2004,Giovannetti2006,Giovannetti2011}.
In comparison to classical strategies, quantum-enhanced measurement offers significant advantages, where a dramatic improvement for the achievable precision can be obtained due to the use of quantum entanglement~\cite{Pezze2009, Hyllus2010, Huang2014}.
Preparing and detecting entangled quantum states are two main challenges to achieve the Heisenberg-limited metrology.
A lot of endeavors had been made to create various kinds of entangled input states, such as spin squeezed states~\cite{ Leroux2010, Gross2010, Riedel2010, Muessel2015}, twin Fock states~\cite{Lucke2011, Zhang2013, Luo2017}, maximally entangled states~\cite{Bollinger1996, Monz2011}, and so on.
However, to detect the output states, single-particle resolved detection is assumed to be necessary, which has so far been the bottleneck in the practical performances.
Moreover, imperfect detection is one of the key obstacles that hamper the improvement of measurement precision via many-body entangled states.

Recently,  an echo protocol was proposed to perform phase estimation near the Heisenberg limit~\cite{Davis2016, Frowis2016}, which does not require single-particle resolved detection.
The input state is generated by the time-evolution under a one-axis twisting Hamiltonian $\hat U=e^{-iH_{\textrm{OAT}}t}$~\cite{Kitagawa1993}, and a reversal evolution $\hat U_R=\hat U^{\dagger}$ is performed on the output state prior to the final population measurement.
The nonlinear dynamics $\hat U_R$ enables Heisenberg-limited precision scaling ($\propto N^{-1}$) under detection noise $\sigma\lesssim\sqrt{N}$, where $N$ is the total particle number.
This kind of nonlinear detection with spin squeezed states~\cite{Hosten2016} and two-mode squeezed vacuum states~\cite{Linnemann2016} has been respectively realized in experiments.
More recently, schemes on interaction-based readout that relax the time-reversal condition ($\hat U_R \neq \hat U^{\dagger}$) have been proposed~\cite{Nolan2017, Mirkhalaf2018}.
These pointed out a new direction of utilizing entangled states for quantum metrology~\cite{Dunningham2002, Macri2016, Szigeti2017, Fang2017, Anders2017, Huang2018, Choi2018, Haine2018}.

Spin cat states, a kind of non-Gaussian entangled states as a superposition of distinct spin coherent states (SCSs), are considered as promising candidates for quantum-enhanced measurement~\cite{Agarwal1997, Gerry1998, Sanders2014, Lau2014, Signoles2014, Huang2015}.
It has been shown that spin cat states with modest entanglement can perform high-precision phase measurement beyond the standard quantum limit (SQL) even under dissipation~\cite{Huang2015}.
However, to perform the interferometry with spin cat states in practice, parity measurement is required so that single-particle resolution should be accessed~\cite{Gerry2010, Zhang2012, Hume2013, Huang2015, LuoC2017}.
The requirement of single-particle detection limits the experimental feasibility of quantum metrology with spin cat states.
Thus, to overcome this barrier, is it possible to replace the parity measurement with interaction-based readout?
Compared with the interaction-based readout scheme with spin squeezed states, will the spin cat states offer better robustness against detection noise?

In this article, we propose to perform the phase estimation with spin cat states via interaction-based readout.
We find that spin cat states have the ability to perform Heisenberg-limited phase measurement and that interaction-based readout is an optimal method to fully exploit this potential ability.
In Sec.~\ref{Sec2}, we give a general framework of many-body quantum interferometry and the phase estimation.
In Sec.~\ref{Sec3}, we analytically obtain the ultimate precision bound for spin cat states.
The ultimate bound is always inversely proportional to the total particle number with a constant depending on the separation of the two superposition SCSs that approaches the Heisenberg limit.
In Sec.~\ref{Sec4}, we describe the procedure of quantum interferometry via interaction-based readout with spin cat states.
Then, the estimated phase precisions via interaction-based readout are numerically calculated and the optimal conditions are given.
Especially, when the estimated phase lies around $\phi \sim \pi/2$, all the spin cat states can saturate their ultimate bounds with suitable interaction-based readout.
Finally, the detailed derivation of how the interaction-based readout can saturate the ultimate precision bounds of spin cat states are analytically shown.
In Sec.~\ref{Sec5}, we analyze the robustness against detection noise within our scheme.
It is demonstrated that, the spin cat states under interaction-based readout are immune to detection noise up to $\sigma \lesssim \tilde{c}_D(\theta) N$ with $\tilde{c}_D(\theta)$ a constant depending on the form of spin cat states.
Compared with the echo twisting schemes, our proposal with spin cat states can be approximately $\sqrt{N}$ times more robust against the excess detection noise.
In addition, the influence of dephasing during the nonlinear evolution in the process of interaction-readout is discussed.
In Sec.~\ref{Sec6}, we briefly summarize our results.

\section{Phase Estimation via Many-body Quantum Interferometry}\label{Sec2}

The most widely used interferometry can be described within a two-mode bosonic system of $N$ particles, such as Ramsey interferometry with ultracold atoms~\cite{Pezze2016}, trapped ions~\cite{Wineland1994, Blatt2008}, and Mach-Zehnder interferometry in optical systems~\cite{Dowling2003}.
In these systems, the system state can be well characterized by the collective spin operators, $\hat J_x = {1\over2} \left(\hat a \hat b^{\dagger} + \hat a^{\dagger} \hat b\right)$, $\hat J_y = {1\over2i} \left(\hat a \hat b^{\dagger} - \hat a^{\dagger} \hat b\right)$, $\hat J_z = {1\over2}\left(\hat b^{\dagger}\hat b - \hat a^{\dagger} \hat a\right)$ with $\hat a $ and $\hat b$ the annihilation operators for particles in mode $|a\rangle$ and mode $|b\rangle$, respectively.
A common quantum interferometry can be divided into three steps.
First, a desired input state $|\psi\rangle_{in}$ is prepared.
Then, the input state evolves under the action of an unknown quantity and accumulates an phase $\phi$ to be measured, i.e., $|\psi(\phi)\rangle_{out}=\hat U(\phi) |\psi\rangle_{in}$.
Finally, a proper sequence of measurement onto the output state $|\psi(\phi)\rangle_{out}$ is implemented to extract the accumulated phase.
Theoretically, for a given phase accumulation process $\hat U(\phi)=e^{-i\phi\hat G}$, the measurement precision of the accumulated phase is constrained by a fundamental limit, the quantum Cram\'{e}r-Rao bound (QCRB)~\cite{Braunstein1994, Huang2014, Pezze2009, Hyllus2010}, which only depends on the specific property of the chosen input state,
\begin{equation}\label{QCRB}
    \Delta \phi \ge \Delta \phi^{Q} \equiv \frac{1}{\sqrt{\mu F^Q}},
\end{equation}
\begin{equation}\label{FQ}
    F^Q = 4\left( \langle \psi'|\psi'\rangle - |\langle \psi'|\psi(\phi)\rangle_{out}|^2\right)=4\Delta^2 \hat G,
\end{equation}
where $\Delta \phi = \sqrt{\langle \phi^2 \rangle - \langle\phi\rangle^2}$ is the standard deviation of the estimated phase, $\mu$ corresponds to the number of trials, $|\psi'\rangle=\text{d}|\psi(\phi)\rangle_{out} / \text{d}\phi$ denotes the derivative and $\Delta^2 \hat G = _{in}\!\!\langle\psi |\hat G^2| \psi\rangle_{in} - _{in}\!\!\langle\psi |\hat G| \psi\rangle_{in}^2$ is the variance of $\hat G$ for the input state.

In realistic scenarios, the frequency shift between the two modes $\omega$ is one of the most widely interesting parameters to be estimated owing to its importance in frequency standards~\cite{Margolis2009}.
Therefore, the generator can be chosen as $\hat G=\hat J_z$, the estimated phase $\phi=\omega t$ and the quantum Fisher information (QFI) becomes $F^Q=4\Delta^2 \hat J_z$.
It is well known that, using an input GHZ state can maximize $F^Q$ to $N^2$, and the corresponding phase measurement precision scales inversely proportional to the total particle number, $\Delta\phi \propto N^{-1}$, attaining the Heisenberg limit.
%
%
%
In the following, we will show that, apart from GHZ state, other spin cat states also have the ability to perform Heisenberg-limited phase estimation.
Further, we will give an experimentally feasible scheme to realize the Heisenberg-limited measurement with spin cat states by means of interaction-based readout.

\section{Ultimate precision bound of spin cat states}\label{Sec3}
Spin cat states are typical kinds of macroscopic superposition of spin coherent states (MSSCS).
Generally, a MSSCS is a superposition of multiple spin coherent states (SCSs)~\cite{Ferrini2008, Ferrini2010, Pawlowski2013}.
Here, we discuss the MSSCS in the form of
\begin{equation}\label{MSSCS}
    |\Psi(\theta, \varphi)\rangle_{\textrm{M}}=\mathcal{N}_{C}(|\theta,\varphi\rangle + |\pi-\theta,\varphi\rangle),
\end{equation}
where $\mathcal{N}_{C}$ is the normalization factor and $\left|\theta,\varphi\right\rangle$ denotes the $N$-particle SCS with
\begin{equation}\label{SCS_Dicke}
    \left|\theta,\varphi\right\rangle=\sum^{J}_{m=-J} c_m(\theta)e^{-i(J+m)\varphi}\left|J,m\right\rangle.
\end{equation}
Here, $c_m(\theta) =\sqrt{\frac{(2J)!}{(J+m)!(J-m)!}}\cos^{J+m}\left({\theta \over 2}\right)\sin^{J-m}\left({\theta \over 2}\right)$, and $\{\left|J,m\right\rangle\}$ represents the Dicke basis with $J={N / 2}$ and $m=-J, -J+1, ..., J-1,  J$.
Without loss of generality, we assume $\varphi=0$,
\begin{eqnarray}\label{MSSCS_Dicke}
    |\Psi(\theta)\rangle_{\textrm{M}}&=&\mathcal{N}_{C}(|\theta,0\rangle + |\pi-\theta,0\rangle) \nonumber\\
    &=& \mathcal{N}_{C}\left(\sum^{J}_{m=-J} \left[c_m(\theta)+c_m(\pi-\theta)\right]\left|J,m\right\rangle\right) \nonumber\\
    &=& \mathcal{N}_{C} \left[\sum^{J}_{m=-J} c_m(\theta)\left(\left|J,m\right\rangle+\left|J,-m\right\rangle\right)\right],
\end{eqnarray}
where the two SCSs have the same azimuthal angle $\varphi=0$ and the polar angles are symmetric about $\theta=\pi/2$.
Since $c_m(\theta)=c_{-m}(\pi-\theta)$, the coefficients of the MSSCS are symmetric about $m=0$.
It is shown that~\cite{Huang2015}, when the two superposition SCSs are orthogonal or quasi-orthogonal, the corresponding MSSCS can be regarded as a spin cat state.
Mathematically, the sufficient condition of spin cat states can be expressed as
\begin{equation}\label{c0}
    \theta \lesssim \theta_c\equiv\sin^{-1}\left\{2\left[\frac{\left((J-1)!\right)^2}{2 (2J)!}\right]^{1\over{2J}}\right\},
\end{equation}
which is derived from the assumption $2 |c_0(\theta)|^2 = \frac{(2J)!}{J! J!} \cos^{2J} \left(\theta\over2\right) \sin^{2J} \left(\theta\over2\right) \lesssim \frac{1}{J^2}$ when $J>1$.
According to Eq.~\eqref{c0},  $\theta_c$ increases as the total particle number $N=2J$ grows.
%
For total particle number $N \ge 40$, one can find $\theta_c \approx 7\pi/20$.
Under this condition, the normalization factor $\mathcal{N}_{C}\approx \frac{1}{\sqrt{2}}$.
Throughout this paper, we will focus on the spin cat states under the conditions of $N \ge 40$ and $\theta \le 7\pi/20$.
We abbreviate the spin cat states $\left|\Psi(\theta)\right\rangle_{\textrm{M}}$ ($N \ge 40$ and $\theta \le 7\pi/20$) to $\left|\Psi(\theta)\right\rangle_{\textrm{CAT}}$ below, i.e.,
\begin{equation}\label{CAT}
    \left|\Psi(\theta)\right\rangle_{\textrm{CAT}}\approx\frac{1}{\sqrt{2}} \left[\sum^{J}_{m=-J} c_m(\theta)\left(\left|J,m\right\rangle+\left|J,-m\right\rangle\right)\right].
\end{equation}

Note that spin cat states can be understood as a superposition of GHZ states with different spin length.
The average of half-population difference $_\textrm{CAT}\langle\Psi(\theta) |\hat J_z| \Psi(\theta)\rangle_\textrm{CAT}=\sum_{m=-J}^{J} m c_m^{*}(\theta) c_m(\theta)=0$ for all spin cat states.
Owing to $_\textrm{CAT}\langle\Psi(\theta) |\hat J_z| \Psi(\theta)\rangle_\textrm{CAT}=0$, the variance of a spin cat state $|\Psi(\theta)\rangle_\textrm{CAT}$ becomes $\Delta^2 \hat J_z=\langle\Psi(\theta) |\hat J_z^2| \Psi(\theta)\rangle_\textrm{CAT} = \sum_{m=-J}^{J} m^2 c_m^{*}(\theta) c_m(\theta)$.
Since the two superposition SCSs are well fragmented for a spin cat state, and the coefficients $c_m(\theta)$ is in a binomial distribution, the variance can be approximately calculated as
\begin{equation}\label{variance}
    \Delta^2 \hat J_z = \sum_{m=-J}^{J} m^2 |c_m(\theta)|^2 \approx \overline M^2,
\end{equation}
where $-\overline M$ and $\overline M$ can be regarded as the center locations of the two peaks.
This assumption is valid for calculating the QFI and perfectly matches the numerical results especially when the total particle number $N$ is large.
Given that $\Delta^2 J_z \approx \overline M^2$, the QFI of a spin cat state can be obtained, i.e., $F^Q_\textrm{CAT} \approx 4\overline M^2$.

\begin{figure}[htb]
\centering
\includegraphics[width=\columnwidth]{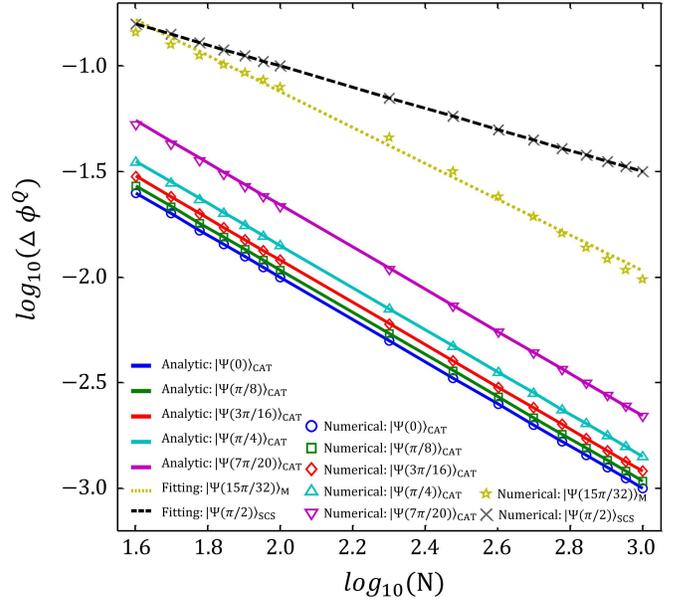}\caption{(Color online) The ultimate precision scalings of different $|\Psi(\theta)\rangle_{\textrm{M}}$. The circles, squares, diamonds, triangles and inverted triangles are the numerical results with $|\Psi(\theta)\rangle_\textrm{CAT}$ for $\theta=0, \pi/8, 3\pi/16, \pi/4, 7\pi/20$, respectively. The solid lines are the analytic results for the spin cat states, which perfectly agree with the numerical ones. The slopes equal to $-1$, indicating that the spin cat states exhibit Heisenberg-limited precision scalings. The pentagrams and crosses denote the numerical results for MSSCS $|\Psi(15\pi/32)\rangle_{\textrm{M}}$ and SCS $|\Psi(\pi/2)\rangle_{\textrm{SCS}}$, which do not satisfy the analytic expression~\eqref{QCRB_M}. Therefore, the dotted and dashed lines are the fitting results for $|\Psi(15\pi/32)\rangle_{\textrm{M}}$ and $|\Psi(\pi/2)\rangle_{\textrm{SCS}}$, where the slopes are $-0.83$ and $-0.5$, respectively.}
\label{Fig1}
\end{figure}

Mathematically, $\overline M$ is a continuous variable, and it can be determined by the equation (see Appendix A),
\begin{equation}\label{Det_M}
    \sqrt{\frac{J+\overline M}{J-\overline M+1}}\tan\left(\frac{\theta}{2}\right)=1.
\end{equation}
Solving Eq.~\eqref{Det_M}, we get $\overline M=J \left( \frac{1-\tan^2\left(\theta/2\right)}{1+\tan^2\left(\theta/2\right)} \right) + \frac{1}{1+\tan^2(\theta/2)}$.
When the total particle number $N=2J$ is large, it can be approximated as
\begin{equation}\label{MM}
    \overline M \approx \left( \frac{1-\tan^2\left(\theta/2\right)}{1+\tan^2\left(\theta/2\right)} \right){N\over2},
\end{equation}
and the QFI of a spin cat state can be written as
\begin{equation}\label{QFI_M}
    F^Q_\textrm{CAT} \approx \left( 1- \frac{2\tan^2\left(\theta/2\right)}{1+\tan^2\left(\theta/2\right)} \right)^2{N^2}.
\end{equation}
Thus, according to the QCRB~\eqref{QCRB}, the ultimate phase precision by a spin cat state is obtained,
\begin{eqnarray}\label{QCRB_M}
    \Delta\phi \ge \Delta\phi^Q_\textrm{CAT} &=&{C(\theta)} N^{-1} \\\nonumber
     &=& \left( 1+ \frac{2\tan^2\left(\theta/2\right)}{1-\tan^2\left(\theta/2\right)} \right) N^{-1}.
\end{eqnarray}
The above analytic result~\eqref{QCRB_M} is verified by numerical calculations, as shown in Fig.~\ref{Fig1}.

From the expression of QCRB~\eqref{QCRB_M}, the achievable precision of the spin cat state $|\Psi(\theta)\rangle_\textrm{CAT}$ follows the Heisenberg scaling multiplied by a coefficient $C(\theta)$ only dependent on $\theta$.
This coefficient $C(\theta)$ grows monotonically as $\theta$ gets larger.
When $\theta=0$, $|\Psi(0)\rangle_\textrm{CAT}$ becomes the GHZ state (an extreme type of spin cat states), its ultimate bound returns to $1/N$.
When $\theta>0$, the ultimate bound becomes $C(\theta)/N$, which is a constant fold higher than the GHZ state.
For example, $\Delta\phi^{Q}=2/N$ for $|\Psi(\pi/4)\rangle_\textrm{CAT}$, for a given $N$, the achievable precision just decreases by half compared to $|\Psi(0)\rangle_\textrm{CAT}$.
This indicates that, the spin cat states with modest $\theta$ may be more experimentally feasible.
They preserve the Heisenberg scaling of precision, and meanwhile are more easily prepared than the maximally entangled states in experiments~\cite{Lee2006, Lee2009, Huang2015, Xing2016, Huang2018}.

It is worth mentioning that, the ultimate bound~\eqref{QCRB_M} is only valid for spin cat states which satisfy the condition~\eqref{c0}.
For other MSSCS $|\Psi(\theta)\rangle_{\textrm{M}}$ in which the overlap between the two SCSs is more significant, the precision scaling no longer remains Heisenberg-limited, but approaches the SQL as $\theta$ gradually increases towards $\pi/2$.

\section{Interaction-based readout with spin cat states}\label{Sec4}
Although we have demonstrated that the spin cat states have the ability to perform Heisenberg-limited parameter estimation, how to saturate the ultimate precision bound and exploit their full potential in practice is a more important problem.
Here, we will propose a practical scheme to implement the Heisenberg-limited quantum metrology with spin cat states by adding a nonlinear dynamics before the population difference measurement.
We will show that, this is an optimal detection scheme which can saturate the ultimate precision bound of the spin cat states.
\subsection{Interaction-based readout}
\begin{figure}[htb]
\centering
\includegraphics[width=\columnwidth]{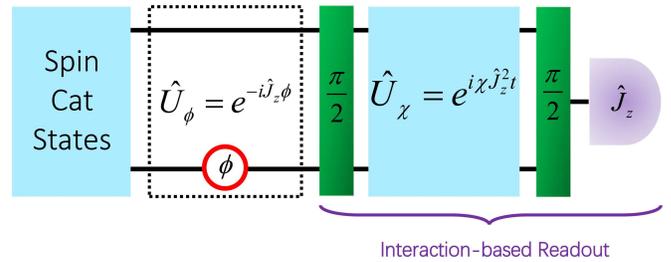}\caption{(Color online) The schematic of the interaction-based readout scheme with input state cat states. First, an input spin cat state is prepared. Then, an estimated phase is accumulated during interrogation. To extract the phase, an interaction-based readout protocol is applied, in which a nonlinear evolution is sandwiched by two $\frac{\pi}{2}$ pulses before the half-population difference measurement.}
\label{Fig2}
\end{figure}

The procedures of our scheme based on interaction-based readout are illustrated as follows, see Fig.~\ref{Fig2}.
First, a suitable input spin cat states $|\Psi(\theta)\rangle_{\textrm{CAT}}$ is prepared.
The spin cat states can be created by several kinds of methods in various quantum systems~\cite{Agarwal1997, Gerry1998, Sanders2014, Lau2014, Signoles2014, Huang2015}.
Particularly in Bose condensed atomic systems, the spin cat states can be generated via nonlinear dynamical evolution~\cite{Ferrini2008, You2003} or deterministically prepared by adiabatic ground state preparation~\cite{Lee2006, Huang2015, Huang2018}.
The parameter $\theta$ for a specific spin cat state is determined by the control of atom-atom interaction~\cite{Gross2010, Riedel2010}.
Then, the input state evolves under the Hamiltonian $H_{0}=\omega \hat J_z$, and the output state $|\Psi(\phi)\rangle_{out}=\hat U(\phi) |\Psi(\theta)\rangle_{\textrm{CAT}}$, where $\hat U(\phi)=e^{-i H_{0} t}=e^{-i\hat J_z\phi}$ with accumulated phase $\phi=\omega t$.
Finally, a sequence interaction-based readout is performed on $|\Psi(\phi)\rangle_{out}$ to extract $\phi$.

Here, the sequence comprises a nonlinear dynamics sandwiched by two $\pi\over2$ pulses prior to the half-population difference measurement.
The final state after the sequence can be written as
\begin{equation}\label{NonDy}
    |\Psi(\phi)\rangle_f = \hat R_x^{\dagger}({\pi\over2}) \hat U_{non}(\chi t) \hat R_x({\pi\over2}) |\Psi(\phi)\rangle_{out},
\end{equation}
where $\hat U_{non}(\chi t)=e^{-iH_{\textrm{OAT}}t}=e^{i\chi\hat J_z^2 t}$ describes the nonlinear evolution with nonlinearity $\chi$, $\hat R_x({\pi\over2})=e^{i{\pi\over2}\hat J_x}$ is a $\pi\over2$ pulse, a rotation about $x$ axis.
%
%
Applying the half-population difference measurement $\hat J_z$ on the final state, one can obtain the expectation and standard deviation of $\hat J_z$,
\begin{equation}\label{Avg_Jz}
    \langle\hat J_z\rangle_f = _{f}\!\langle\Psi(\phi)| \hat J_z |\Psi(\phi)\rangle_{f},
\end{equation}
\begin{equation}\label{Delta_Jz}
    \left(\Delta \hat J_z\right)_f = \sqrt{_{f}\langle\Psi(\phi)| \hat J_z^2 |\Psi(\phi)\rangle_{f} - \left(_{f}\langle\Psi(\phi)| \hat J_z |\Psi(\phi)\rangle_{f}\right)^2}.
\end{equation}
Therefore, the estimated phase precision is given according to the error propagation formula,
\begin{equation}\label{Delta_phi}
    \Delta \phi = \frac{\left(\Delta \hat J_z\right)_f}{\left|\partial \langle\hat J_z\rangle_f / \partial \phi \right|}.
\end{equation}

\subsection{Numerical results}
\begin{figure*}[htb]
\centering
\includegraphics[scale=0.5]{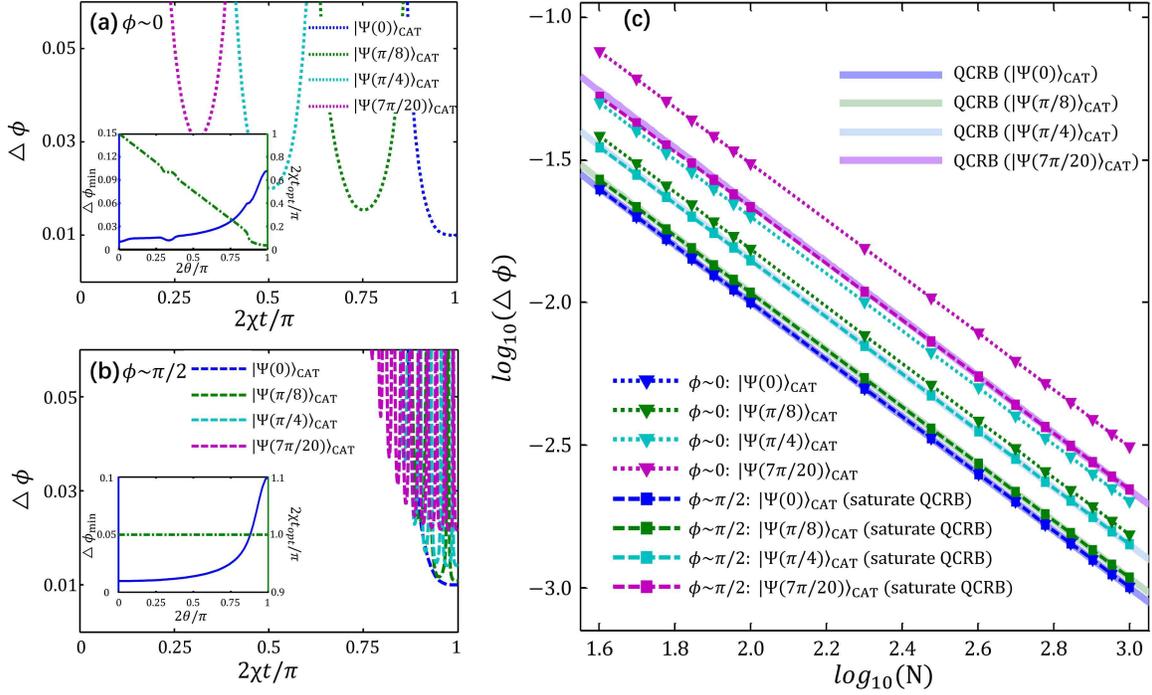}\caption{(Color online) The phase precision obtained with spin cat states via interaction-based readout for the estimated phase in the vicinity of (a) $\phi\sim0$ and (b) $\phi\sim\pi/2$. The standard derivations of the phase $\Delta \phi$ versus the nonlinear evolution $\chi t$ for spin cat states $|\Psi(0)\rangle_{\textrm{CAT}}$ (blue), $|\Psi(\pi/8)\rangle_{\textrm{CAT}}$ (green), $|\Psi(\pi/4)\rangle_{\textrm{CAT}}$ (cyan) and $|\Psi(7\pi/20)\rangle_{\textrm{CAT}}$ (magenta) are shown. The minimum phase precision $\Delta \phi_\textrm{min}$ is obtained by optimizing the nonlinear evolution $\chi t$. In the insets, $\Delta \phi_\textrm{min}$  and the corresponding optimal evolution time $t_{opt}$ for different MSSCS $|\Psi(\theta)\rangle_{\textrm{M}}$ are given. Here, the total particle number $N=100$. (c) The log-log scaling of the minimum phase precision versus the total particle number for spin cat states $|\Psi(0)\rangle_{\textrm{CAT}}$ (blue), $|\Psi(\pi/8)\rangle_{\textrm{CAT}}$ (green), $|\Psi(\pi/4)\rangle_{\textrm{CAT}}$ (cyan) and $|\Psi(7\pi/20)\rangle_{\textrm{CAT}}$ (magenta). The dotted lines with inverted triangles and the dashed lines with squares correspond to the cases of $\phi\sim0$ and $\phi\sim\pi/2$, respectively.}
\label{Fig3}
\end{figure*}

The measurement precision via interaction-based readout with spin cat states are shown in Fig.~\ref{Fig3}.
We first consider the accumulated phase around $\phi=0$, and plot the precision dependence on the nonlinear evolution, see Fig.~\ref{Fig3}~(a).
The optimal nonlinear evolution $\chi t$ changes with different input spin cat states $|\Psi(\theta)\rangle_{\textrm{CAT}}$.
For spin cat states with larger $\theta$, despite the estimated phase precision $\Delta \phi_{\textrm{min}}$ becomes a bit worse, but the required optimal nonlinear evolution $\chi t$ is getting smaller.
Given a fixed nonlinearity $\chi$, the optimal nonlinear evolution time $t_{opt}$ decreases with $\theta$, see the inset of Fig.~\ref{Fig3}~(a).
We choose four typical spin cat states $|\Psi(0)\rangle_{\textrm{CAT}}$, $|\Psi(\pi/8)\rangle_{\textrm{CAT}}$, $|\Psi(\pi/4)\rangle_{\textrm{CAT}}$ and $|\Psi(7\pi/20)\rangle_{\textrm{CAT}}$ and evaluate their precision scaling versus total particle number, see Fig.~\ref{Fig3}~(c).
It is shown that, the spin cat states with interaction-based readout still preserve the Heisenberg-limited scaling when $\phi \sim 0$.
Although there exists a shift from the QCRB for spin cat states with large $\theta$, the Heisenberg scaling $\Delta\phi_{\textrm{min}} \propto 1/N$ enables the high-precision measurement when the total particle number is large.

Further, we also consider the accumulated phase around $\phi=\pi/2$, and plot the precision dependence on the nonlinear evolution, see Fig.~\ref{Fig3}~(b).
Differently from the case of $\phi \sim 0$, the optimal nonlinear evolution $\chi t = \pi/2$ for all input spin cat states $|\Psi(\theta)\rangle_{\textrm{CAT}}$ when $\phi \sim \pi/2$, see Fig.~\ref{Fig3}~(b).
The precision scaling versus total particle number saturate the QCRB~\eqref{QCRB_M}, which indicates that the interaction-based readout is an optimal scheme to attain the ultimate bound of the spin cat states, see Fig.~\ref{Fig3}~(c).

The measurement precision can also be estimated via classical Fisher information (CFI).
We numerically find that, the minimum standard deviations in the above scenarios are the same with the results calculated by CFI.

Both scenarios are useful in practical parameter estimation.
When the parameter is very tiny, the accumulated phase may be around $\phi=0$, the spin cat states with modest $\theta$ are beneficial.
For example, the optimal nonlinear evolution of $|\Psi(\pi/4)\rangle_{\textrm{CAT}}$ is $\chi t =\pi/4$, which is only half of the one for the GHZ state.
Meanwhile, the corresponding precision scaling is still $\Delta\phi \propto 1/N$.
On the other hand, for relatively large parameters and the interrogation time can be varied so that the accumulated phase lies around $\phi=\pi/2$, the interaction-based readout can saturate the ultimate bound only if the nonlinear evolution can be tuned to $\chi t =\pi/2$.

\begin{figure*}[htb]
\centering
\includegraphics[scale=0.5]{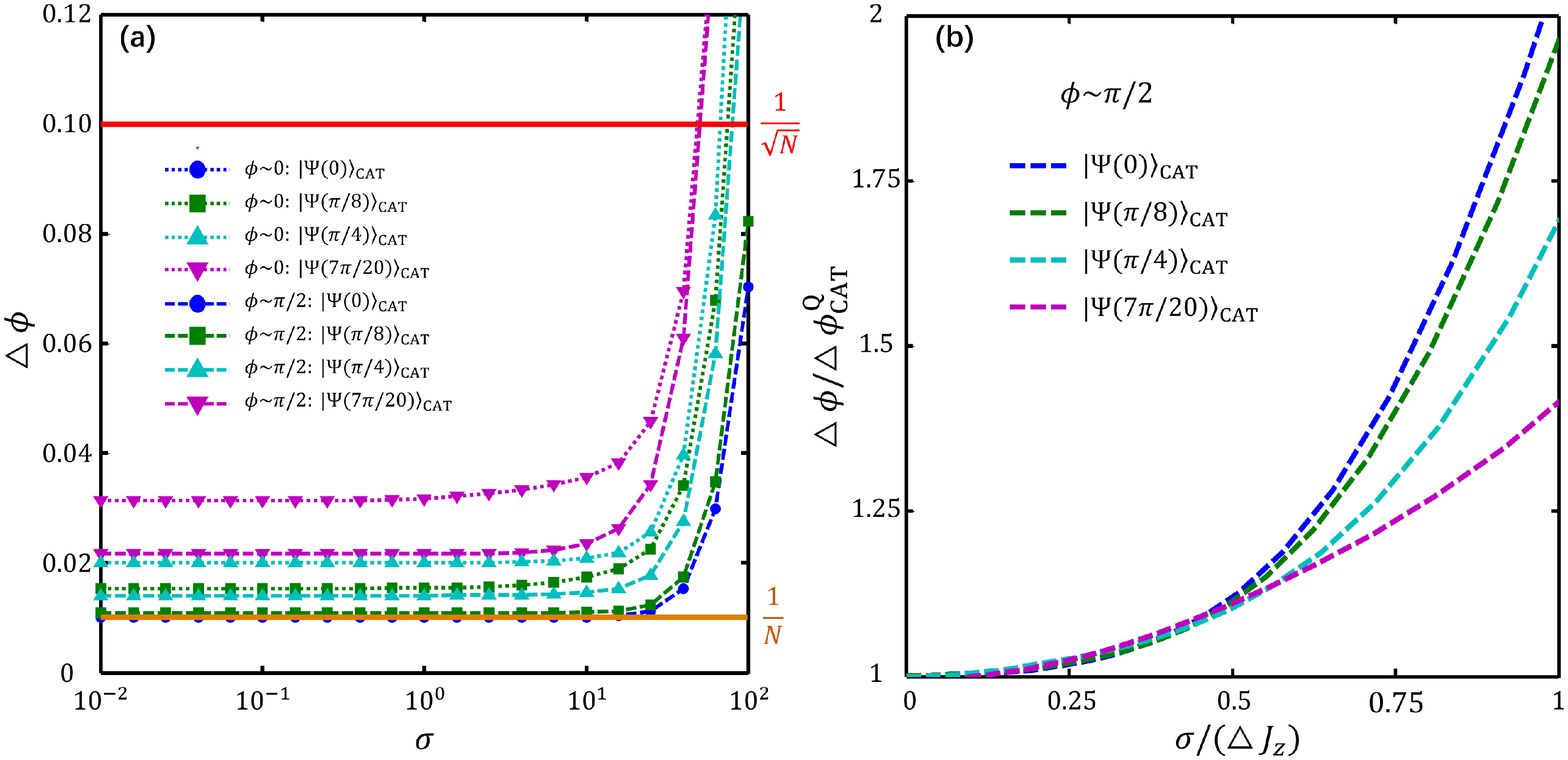}\caption{(Color online) (a) The optimal phase uncertainty under optimal control  against detection noise $\sigma$ with spin cat states $|\Psi(0)\rangle_{\textrm{CAT}}$ (blue), $|\Psi(\pi/8)\rangle_{\textrm{CAT}}$ (green), $|\Psi(\pi/4)\rangle_{\textrm{CAT}}$ (cyan) and $|\Psi(7\pi/20)\rangle_{\textrm{CAT}}$ (magenta). The dotted and dashed lines correspond to the cases of $\phi\sim0$ and $\phi\sim\pi/2$, respectively. Here, the total particle number is $N=100$. (b) The relative phase standard deviation $\Delta\phi/\Delta\phi_{CAT}^Q$ versus $\sigma/(\Delta \hat{J}_z)$ for spin cat states $|\Psi(\theta)\rangle_{\textrm{CAT}}$. Here, $\Delta\phi_{CAT}^Q\approx(2\overline{M})^{-1}$, $\Delta \hat{J}_z \approx \overline{M}$, and the relative phase standard deviation begins to blow up when $\sigma\approx0.5\overline{M}\approx N/2C(\theta)=\tilde{c}_D(\theta) N$.}
\label{Fig4}
\end{figure*}

\subsection{Analytical analysis}
For spin cat states via interaction-based readout, the corresponding measurement precision can be analyzed analytically for some specific cases.
We will show how the interaction-based readout with $\chi t=\pi/2$ can saturate the ultimate precision bound for spin cat states when the estimated phase is around $\phi=\pi/2$.
Then, we will also illustrate the reason why the interaction-based readout with spin cat states for $\phi=0$ and $\phi=\pi/2$ make a difference.

Consider an input state,
\begin{equation}\label{Psi_in_gen}
    |\Psi_{in}\rangle=\sum_{k=-J}^{J} a_k |J,k\rangle,
\end{equation}
which is symmetric with respect to exchange of two modes, where $a_{k}=a_{-k}$ and $J=N/2$ is an even number.

According to Eq.~\eqref{NonDy}, the final state before observable measurement can be expressed as,
\begin{eqnarray}\label{Psi_f_gen}
    |\Psi(\phi)\rangle_f &=& e^{-i\frac{\pi}{2}\hat J_x} e^{i\chi t\hat J_z^2} e^{i\frac{\pi}{2}\hat J_x} e^{-i \phi \hat J_z}|\Psi_{in}\rangle, \nonumber\\
    &=& \sum_{k=-J}^{J} a_k \left(e^{-i\frac{\pi}{2}\hat J_x} e^{i\chi t\hat J_z^2} e^{i\frac{\pi}{2}\hat J_x} e^{-i \phi \hat J_z}\right)|J,k\rangle. \nonumber\\
    &&
\end{eqnarray}
When $\chi t=\pi/2$, Eq.~\eqref{Psi_f_gen} becomes,
\begin{equation}\label{Psi_f_gen2}
    |\Psi(\phi)\rangle_f=\sum_{k=-J}^{J} a_k \left(e^{-i\frac{\pi}{2}\hat J_x} e^{i\frac{\pi}{2} J_z^2} e^{i\frac{\pi}{2}\hat J_x} e^{-i \phi \hat J_z}\right)|J,k\rangle.
\end{equation}
Since $a_{k}=a_{-k}$,
\begin{eqnarray}\label{Psi_f_gen3}
    && |\Psi(\phi)\rangle_f= a_0\left(e^{-i\frac{\pi}{2}\hat J_x} e^{i\frac{\pi}{2} J_z^2} e^{i\frac{\pi}{2}\hat J_x} e^{-i \phi \hat J_z}\right)|J,0\rangle \nonumber\\
    &+& \sum_{k=1}^{J} a_k \left(e^{-i\frac{\pi}{2}\hat J_x} e^{i\frac{\pi}{2} J_z^2} e^{i\frac{\pi}{2}\hat J_x} e^{-i \phi \hat J_z}\right)\left(|J,k\rangle + |J,-k\rangle\right). \nonumber\\
    &&
\end{eqnarray}
One can prove that (see Appendix B),
\begin{eqnarray}\label{Psi_nonDy}
    && e^{-i\frac{\pi}{2}\hat J_x} e^{i\frac{\pi}{2} J_z^2} e^{i\frac{\pi}{2}\hat J_x} e^{-i \phi \hat J_z}\left(|J,k\rangle + |J,-k\rangle\right) \nonumber\\
    &=&\left[\cos(k\phi)+\sin(k\phi)\right]\left(i\right)^{(J-k)^2}|J,k\rangle \nonumber\\
    &+&\left[\cos(k\phi)-\sin(k\phi)\right]\left(i\right)^{(J-k)^2}|J,-k\rangle
\end{eqnarray}
Using Eq.~\eqref{Psi_nonDy}, we can get
\begin{eqnarray}\label{Psi_f_gen4}
    |\Psi(\phi)\rangle_f=\sum_{k=-J}^{J} a_k &&\left[\cos(k\phi)+ (-1)^{J-k}\sin(k\phi)\right] \nonumber\\
    && \left(i\right)^{(J-k)^2}|J,k\rangle.
\end{eqnarray}
Then, the conditional probability of obtaining measurement result of $k$ can also be obtained,
\begin{eqnarray}\label{ConPro}
    P(k|\phi)&=&|a_k|^2 \left[\cos(k\phi)+(-1)^{J-k}\sin(k\phi)\right]^2 \nonumber\\
            &=&|a_k|^2 \left[1+(-1)^{J-k}\sin(2k\phi)\right].
\end{eqnarray}
Thus, the CFI can be calculated,
\begin{eqnarray}\label{CFI}
    F^C(\phi)&=&\sum_{k}\frac{1}{P(k|\phi)}\left(\frac{\partial P(k|\phi)}{\partial \phi}\right)^2 \nonumber\\
            &=& \sum_{k}\frac{4|a_k|^4 k^2\cos^2(2k\phi)}{|a_k|^2[1+(-1)^{J-k}\sin(2k\phi)]},
\end{eqnarray}
and the corresponding Cram\'{e}r-Rao bound is $\Delta\phi^C=1/{\sqrt{F^C(\phi)}}$.
One can easily find that, when $\phi=\pi/2$, $F^C=\sum_k 4 |a_k|^2 k^2$.
For spin cat states, $F^C=4\sum_{k=-J}^{J} k^2 |c_k(\theta)|^2 =F^Q\approx 4\overline M^2$, $\Delta\phi^C=\Delta\phi^Q$.
This indicates that, the ultimate precision bound (i.e., QCRB) can be saturated via the CFI~\eqref{CFI}.
To obtain the CFI, one need to estimate the full probability distribution of the final state~\cite{Szigeti2017, Haine2018, Mirkhalaf2018}.

In our scheme, one can also saturate the ultimate precision bound by measuring the expectation of half-population difference and using the error propagation formula.
The half-population difference of the final state can be written explicitly,
\begin{eqnarray}\label{Avg_Jz_gen}
    \langle\hat J_z\rangle_f &=& _{f}\langle\Psi(\phi)| \hat J_z |\Psi(\phi)\rangle_{f} \nonumber\\
    &=& \sum_{k=-J}^{J} k |a_k|^2 \left[\cos(k\phi)+ (-1)^{J-k}\sin(k\phi)\right]^2 \nonumber\\
    &=& \sum_{k=-J}^{J} k |a_k|^2 \left[1+ (-1)^{J-k}\sin(2k\phi)\right] \nonumber\\
    &=& \sum_{k=-J}^{J} k |a_k|^2 (-1)^{J-k}\sin(2k\phi),
\end{eqnarray}
and its derivative with respect to $\phi$ reads as,
\begin{equation}\label{Derivative_gen}
    \frac{d\langle\hat J_z\rangle_f}{d\phi} = \sum_{k=-J}^{J} 2k^2 |a_k|^2 (-1)^{J-k}\cos(2k\phi).
\end{equation}
Correspondingly, the standard deviation of half-population difference is
\begin{eqnarray}\label{Delta_Jz_gen}
    \left(\Delta \hat J_z\right)_f &=& \sqrt{_{f}\langle\Psi(\phi)| \hat J_z^2 |\Psi(\phi)\rangle_{f} - \left(_{f}\langle\Psi(\phi)| \hat J_z |\Psi(\phi)\rangle_{f}\right)^2} \nonumber\\
    &=& \{\sum_{k=-J}^{J} k^2 |a_k|^2 \left[1+ (-1)^{J-k}\sin(2k\phi)\right] \nonumber\\
    &-& \left[\sum_{k=-J}^{J} k |a_k|^2 (-1)^{J-k}\sin(2k\phi)\right]^2\}^{1/2} \nonumber\\
    &=& \!\sqrt{\sum_{k=-J}^{J} \!k^2 |a_k|^2 \!-\! \left[\sum_{k=-J}^{J} k |a_k|^2 (\!-\!1)^{J-k}\sin(2k\phi)\!\right]^2}. \nonumber\\
    &&
\end{eqnarray}
Finally, we can obtain the phase measurement precision via Eqs.~\eqref{Derivative_gen} and~\eqref{Delta_Jz_gen},
\begin{equation}\label{Delta_phi_gen}
    \Delta \phi = \!\frac{\sqrt{\sum_{k=-J}^{J} k^2 |a_k|^2 \!-\! \left[\!\sum_{k=-J}^{J} k |a_k|^2 (-1)^{J-k}\sin(2k\phi)\!\right]^2}}{\left|\sum_{k=-J}^{J} 2k^2 |a_k|^2 (-1)^{J-k}\cos(2k\phi)\right|}.
\end{equation}

When $\phi=0$, $\sin(2k\phi)=0$ and $\cos(2k\phi)=1$, the phase measurement precision becomes
\begin{equation}\label{Delta_phi_gen_0}
    \Delta \phi|_{\phi=0} = \frac{\sqrt{\sum_{k=-J}^{J} k^2 |a_k|^2}}{\left|\sum_{k=-J}^{J} 2k^2 |a_k|^2 (-1)^{J-k}\right|}.
\end{equation}
When $\phi=\pi/2$, $\sin(2k\phi)=0$ and $(-1)^{J-k}\cos(2k\phi)=1$, the phase measurement precision can be simplified as
\begin{equation}\label{Delta_phi_gen_halfpi}
    \Delta \phi|_{\phi=\pi/2} = \frac{\sqrt{\sum_{k=-J}^{J} k^2 |a_k|^2}}{\left|\sum_{k=-J}^{J} 2k^2 |a_k|^2 \right|}.
\end{equation}

For a spin cat state~\eqref{CAT}, according to Eqs.~\eqref{variance},~\eqref{MM},~\eqref{QCRB_M},~\eqref{Delta_phi_gen_0} and~\eqref{Delta_phi_gen_halfpi}, we get the phase measurement precision
\begin{equation}\label{Delta_phi_CAT_0}
    \Delta \phi |_{\phi=0}= \frac{\overline M}{\left|\sum_{m=-N/2}^{N/2} 2k^2 |c_m(\theta)|^2 (-1)^{N/2-m}\right|},
\end{equation}
and
\begin{eqnarray}\label{Delta_phi_CAT_halfpi}
    \Delta \phi |_{\phi=\pi/2} &=& \frac{\overline M}{\left|\sum_{m=-N/2}^{N/2} 2k^2 |c_m(\theta)|^2\right|} \nonumber\\
    &=& \frac{\overline M}{2\overline M^2} \nonumber\\
    &=& C(\theta) N^{-1}.
\end{eqnarray}
From Eq.~\eqref{Delta_phi_CAT_halfpi}, it is obvious that the interaction-based readout with $\chi t=\pi/2$ attains the ultimate precision bound $\Delta_{CAT}^Q$ of spin cat states when $\phi=\pi/2$.
Comparing with Eq.~\eqref{Delta_phi_CAT_0} and Eq.~\eqref{Delta_phi_CAT_halfpi}, we find that, for interaction-based readout with $\chi t=\pi/2$, $\Delta \phi |_{\phi=0} \ge \Delta \phi |_{\phi=\pi/2}$ since the factor $(-1)^{N/2-m}$ in Eq.~\eqref{Delta_phi_CAT_0} decreases the sensitivity $d\langle\hat J_z\rangle_f/{d\phi}$ when even and odd $m$ coexist.
This also indicates that the interaction-based readout with $\chi t=\pi/2$ may not be the optimal choice for $\phi=0$ with spin cat states.
For $\phi=0$, it is hard to analyze analytically, so we can only obtain the optimal conditions for different spin cat states numerically, which has been shown in the above subsection B.

\section{Robustness against Imperfections}\label{Sec5}
Finally, we investigate the robustness of the interaction-based readout scheme.
In realistic experiments, there are many imperfections that limit the final estimation precision.
Here, we discuss two main sources: the detection noise of the measurement and the dephasing during the nonlinear evolution of the interaction-based readout.

\subsection{Influences of detection noise}
Ideally, the half-population difference measurement onto the final state can be rewritten as $\langle\hat J_z\rangle_f = \sum_{m=-N/2}^{N/2} P_m(\phi) m$, where $P_m(\phi)$ is the measured probability of the final state projecting onto the basis $|J,m\rangle$.
For an inefficient detector with Gaussian detection noise~\cite{Nolan2017,Mirkhalaf2018,Haine2018}, the half-population difference measurement becomes
\begin{equation}\label{GuassianNoise}
    \langle\hat J_z\rangle_f^{\sigma} =  \sum_{m=-N/2}^{N/2} P_m(\phi|\sigma) m,
\end{equation}
with
\begin{equation}\label{GuassianPro}
    P_m(\phi|\sigma)=\sum_{n=-N/2}^{N/2} A_n e^{-(m-n)^2/2\sigma^2} P_n(\phi),
\end{equation}
the conditional probability depending on the detection noise $\sigma$. Here, $A_n$ is a normalization factor.

In Fig.~\ref{Fig4}~(a), we plot the optimal standard deviation $\Delta\phi$ versus the detection noise $\sigma$ with different input spin cat states under the conditions of $\phi\sim0$ and $\phi\sim\pi/2$.
First, we should mention that, the results via CFI when $\phi\sim0$ and $\phi\sim\pi/2$ are the same as the ones using the error propagation formula.
The standard deviation $\Delta\phi$ stays unchanged when $\sigma\le\sqrt{N}$ and starts to blow up as $\sigma$ becomes large enough.
It is obvious that, the spin cat states with smaller $\theta$ are more robust against the detection noise.
To analyze more clearly, for $\phi\sim\pi/2$, we show the relation of $\Delta\phi/\Delta\phi_{CAT}^Q$ versus $\sigma/(\Delta \hat J_z)$, where $\Delta\phi_{CAT}^Q \approx 1/(2\overline M)\approx C(\theta)/N$, $\Delta \hat J_z\approx \overline M$ are the ultimate precision bound and the standard deviation for a spin cat state $|\Psi(\theta)\rangle_{\textrm{CAT}}$, respectively.
Interestingly, all the spin cat states have similar scaling when $\sigma/(\Delta \hat J_z)\lesssim0.5$, as shown in Fig.~\ref{Fig4}~(b).
When $\sigma/(\Delta \hat J_z)>0.5$, the phase uncertainties start to increase rapidly.
The critical point of the detection noise can be expressed as
\begin{equation}\label{DetNoise}
    \sigma_c\approx0.5\Delta \hat J_z\approx 0.5\overline M\approx N/2C(\theta)\equiv \tilde{c}_D(\theta)N.
\end{equation}
Thus, the interaction-based readout with spin cat state is robust against detection noise up to $\sigma_c\propto N$, which is agree with the results in Ref.~\cite{Fang2017}.
Compared with the echo twisting schemes~\cite{Davis2016, Frowis2016}, our proposal will be much more robust against excess detection imperfection when $N$ is relatively large.
In addition, for a spin cat state $|\Psi(\theta)\rangle_{\textrm{CAT}}$ with smaller $\theta$, $\tilde{c}_D(\theta)=1/2C(\theta)$ is larger.
This explains why $|\Psi(\theta)\rangle_{\textrm{CAT}}$ with smaller $\theta$ is more robust in our scheme.

\subsection{Influences of dephasing during interaction-based readout}
Another imperfection may come from environment effects during the process of interaction-based readout.
Here, we consider $\phi\sim0$ and the interrogation duration is shorter than the duration of interaction-based readout. %
Thus the interaction-based readout may suffer from correlated dephasing.
The process can be described by a Lindblad master equation~\cite{Dorner2012},
\begin{equation}\label{Lindblad}
    \frac{d\rho}{dt}=i\left[\chi\hat J_z^2,\rho\right]+\gamma\left(\hat J_z\rho\hat J_z-\frac{1}{2}\hat J_z^2\rho-\frac{1}{2}\rho\hat J_z^2\right),
\end{equation}
where $\gamma$ denotes the dephasing rate and $\rho$ is the density matrix of the evolved state.
The initial density matrix is $\rho(0)=|\tilde{\Psi}\rangle\langle\tilde{\Psi}|$ with $|\tilde{\Psi}\rangle=\hat R_x({\pi\over2}) \hat U(\phi) |\Psi(\theta)\rangle_{\textrm{CAT}}$.

\begin{figure}[htb]
\centering
\includegraphics[width=\columnwidth]{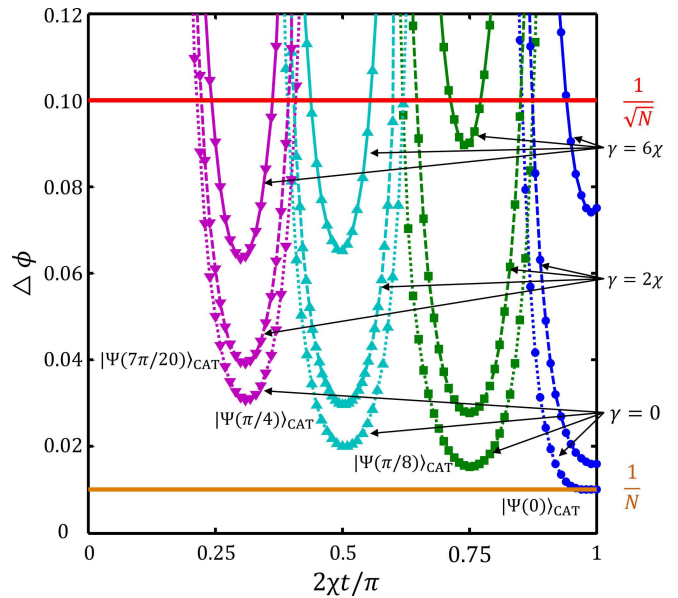}\caption{(Color online) The phase precision $\Delta \phi$ versus the nonlinear evolution $\chi t$ for spin cat states $|\Psi(0)\rangle_{\textrm{CAT}}$ (blue), $|\Psi(\pi/8)\rangle_{\textrm{CAT}}$ (green), $|\Psi(\pi/4)\rangle_{\textrm{CAT}}$ (cyan) and $|\Psi(7\pi/20)\rangle_{\textrm{CAT}}$ (magenta) under dephasing. Here, we consider the accumulated phase is $\phi\sim0$ and after that, the interaction-based readout process is suffered from dephasing with $\gamma$ the dephasing rate. The dotted, dashed and solid lines denote the cases of $\gamma=0$, $\gamma=2\chi$, and $\gamma=6\chi$, respectively. Here, the total particle number is $N=100$.}
\label{Fig5}
\end{figure}

In Fig.~\ref{Fig5}, the effects of dephasing on the estimated phase precision for spin cat states are shown.
First, spin cat states are robust against the dephasing during the interaction-based readout.
The measurement precision can be still beyond SQL when $\gamma$ is large.
Second, the precision of spin cat states with larger $\theta$ degrades more slowly when $\gamma$ becomes severe since the corresponding optimal evolution time is shorter.
Therefore, it is more feasible to use spin cat states with modest $\theta$ via interaction-based readout when the estimated phase is near 0.

\section{Summary}\label{Sec6}

In summary, we have investigated the metrological performances of spin cat states and proposed to implement interaction-based readout to make full use of spin cat states for quantum phase estimation.
We analytically show that spin cat states have the ability to perform Heisenberg-limited measurement, whose standard derivations of the estimated phase are always inversely proportional to the total particle number.
We find that interaction-based readout is one of the optimal methods for spin cat states to perform Heisenberg-limited measurement.
When the estimated phase $\phi$ is around 0, the spin cat states with modest entanglement are beneficial since their optimal nonlinear evolution of the interaction-based readout $\chi t$ is much smaller than $\pi/2$.
However, when the estimated phase $\phi$ lies near $\pi/2$, the interaction-based readout with spin cat states can always saturate the ultimate precision bound if the nonlinear evolution can be tuned to $\chi t =\pi/2$.
The detailed derivation of how the interaction-based readout can saturate the ultimate precision bounds of spin cat states are analytically given.

Moreover, the interaction-based readout with spin cat states is robust against detection noise and it does not require single-particle resolution detectors.
Compared with the twisting echo schemes, our proposal can be immune against detection noise up to $\sigma \sim \tilde{c}_D(\theta) N$, which is much more robust.
Besides, the influences of other imperfect effects such as dephasing during interaction-based readout are also discussed.
Our study on quantum phase estimation with spin cat states via the interaction-based readout may open up a feasible way to achieve Heisenberg-limited quantum metrology with non-Gaussian entangled states.

\section*{Acknowledgements}
We thank Prof. Simon Haine for his helpful discussion. This work is supported by the National Natural Science Foundation of China (NNSFC) under Grants No. 11574405 and No. 11704420. J. H. is partially supported by the National Postdoctoral Program for Innovative Talents of China (BX201600198).

\setcounter{equation}{0}
\renewcommand{\theequation}{A\arabic{equation}}

\section*{APPENDIX A: DETERMINATION OF $\mathbf{\overline M}$ FOR SPIN CAT STATES}
For a spin cat state, since $-\overline M$ and $\overline M$ can be interpreted as the center locations of the two peaks of the spin coherent states, the value of $\overline M$ can be determined by the maximum of coefficient $c_m(\theta)$.
Without loss of generality, we assume $\overline M>0$.
The difference between the nearest two coefficients can be calculated as
\begin{eqnarray}\label{Dif_Coef}
    c_m(\theta)\!&-&\!c_{m\!-\!1}(\theta)=\!\sqrt{\!\frac{(2J)!}{(\!J\!+\!m)!(\!J\!-\!m)!}}\cos^{\!J+\!m}\!\left({\theta \over 2}\right)\sin^{\!J-\!m}\!\left({\theta \over 2}\right) \nonumber\\
    &-&\!\sqrt{\!\frac{(2J)!}{(\!J\!+\!m\!-\!1)!(\!J\!-\!m\!+\!1)!}}\cos^{\!J+\!m-\!1}\!\left({\theta \over 2}\right)\sin^{\!J\!-\!m\!+\!1}\!\left({\theta \over 2}\right)\nonumber\\
    &=& c_m(\theta)\left[1-\sqrt{\frac{J+m}{J-m+1}}\tan\left(\frac{\theta}{2}\right)\right].
\end{eqnarray}
For $m\le\overline M$, $c_m(\theta)\!-\!c_{m\!-\!1}(\theta)\ge0$, while for $m\ge\overline M$, $c_m(\theta)\!-\!c_{m\!-\!1}(\theta)\le0$, therefore when $m=\overline M$, it should be $c_m(\theta)\!-\!c_{m\!-\!1}(\theta)=0$.
Thus,
\begin{equation}\label{A2}
    c_{\overline M}(\theta)\!-\!c_{\overline M\!-\!1}(\theta)=c_{\overline M}(\theta)\left[1\!-\!\sqrt{\frac{J+\overline M}{\!J\!-\!\overline M\!+\!1}}\tan\left(\frac{\theta}{2}\right)\right]=0,
\end{equation}
and since the coefficients $c_m(\theta)$ are all real positive numbers, we can deduce that,
\begin{equation}\label{A3}
    \sqrt{\frac{J+\overline M}{J-\overline M+1}}\tan\left(\frac{\theta}{2}\right)=1.
\end{equation}
Solving Eq.~\eqref{A3}, we can find that,
\begin{equation}\label{A4}
    \overline M=J \left( \frac{1-\tan^2\left(\theta/2\right)}{1+\tan^2\left(\theta/2\right)} \right) + \frac{1}{1+\tan^2(\theta/2)}.
\end{equation}
Since $J\gg1$ and $1/[1+\tan^2\left(\theta/2\right)]<1$ can be neglected,
\begin{equation}\label{A5}
    \overline M \approx J\left( \frac{1-\tan^2\left(\theta/2\right)}{1+\tan^2\left(\theta/2\right)} \right).
\end{equation}

\section*{APPENDIX B: THE EFFECT OF INTERACTION-BASED READOUT ($\chi t=\pi/2$)}
Here, we give the proof of Eq.~\eqref{Psi_nonDy} in the main text.
The Dicke basis,
\begin{equation}\label{B1}
    |J,-J\rangle \equiv \prod_{l=1}^{2J}|\downarrow\rangle_{l},
\end{equation}
\begin{equation}\label{B2}
    |J, J\rangle \equiv \prod_{l=1}^{2J}|\uparrow\rangle_{l},
\end{equation}
and
\begin{equation}\label{B3}
    |J,k\rangle \equiv \prod_{l=1}^{J+k}|\uparrow\rangle_{l}\prod_{l=J+k+1}^{2J}|\downarrow\rangle_{l},
\end{equation}
for $-J<k<J$.
First,
\begin{eqnarray}\label{B4}
    && e^{i\frac{\pi}{2}\hat J_x} e^{-i \phi \hat J_z}\left(|J, J\rangle +|J, -J\rangle\right)\nonumber\\
    &=& e^{i\frac{\pi}{4}\left(\sum_{l=1}^{2J}\hat \sigma_x^{(l)}\right)} e^{-i\frac{\phi}{2}\left(\sum_{l=1}^{2J}\hat \sigma_z^{(l)}\right)} \left(\prod_{l=1}^{2J}|\downarrow\rangle_{l} + \prod_{l=1}^{2J}|\uparrow\rangle_{l}\right)\nonumber\\
    &=& e^{i\frac{\pi}{4}\left(\sum_{l=1}^{2J}\hat \sigma_x^{(l)}\right)} \left(\prod_{l=1}^{2J} e^{i\frac{\phi}{2}}|\downarrow\rangle_{l} + \prod_{l=1}^{2J}e^{-i\frac{\phi}{2}}|\uparrow\rangle_{l}\right)\nonumber\\
    &=& e^{i\phi J}\prod_{l=1}^{2J}\left(\frac{1}{\sqrt{2}}|\downarrow\rangle_{l}+\frac{i}{\sqrt{2}}|\uparrow\rangle_{l}\right) \nonumber\\
    &+& e^{-i\phi J} \prod_{l=1}^{2J}\left(\frac{i}{\sqrt{2}}|\downarrow\rangle_{l}+\frac{1}{\sqrt{2}}|\uparrow\rangle_{l}\right)\nonumber\\
    &=& \sum_{m=-J}^{J} \left[\frac{\sqrt{C_{2J}^{J-m}}}{2^J}(i)^{J+m}e^{i\phi J} |J,m\rangle\right] \nonumber\\
    &+& \sum_{m=-J}^{J} \left[\frac{\sqrt{C_{2J}^{J+m}}}{2^J}(i)^{J-m}e^{-i\phi J}|J,m\rangle\right] \nonumber\\
    &=& \sum_{m=-J}^{J} \frac{\sqrt{C_{2J}^{J-m}}}{2^J}\left[(i)^{J+m}e^{i\phi J}+(i)^{J-m}e^{-i\phi J}\right]|J,m\rangle \nonumber\\
    &&
\end{eqnarray}
Here, $C_{y}^{x}=\frac{y!}{x!(y-x)!}$ is the binomial coefficient. Then,
\begin{eqnarray}\label{B5}
    && e^{i \frac{\pi}{2} \hat J_z^2} e^{i\frac{\pi}{2}\hat J_x} e^{-i \phi \hat J_z}\left(|J, J\rangle +|J, -J\rangle\right)\nonumber\\
    &=& \sum_{m=-J}^{J} \frac{\sqrt{C_{2J}^{J-m}}}{2^J}(i)^{m^2} \left[(i)^{J+m}e^{i\phi J}+(i)^{J-m}e^{-i\phi J}\right]|J,m\rangle.\nonumber\\
    &&
\end{eqnarray}
Considering the cases of even and odd $m$ respectively, we surprisingly find that,
\begin{eqnarray}\label{B6}
    && (i)^{m^2} \left[(i)^{J+m}e^{i\phi J}+(i)^{J-m}e^{-i\phi J}\right]\nonumber\\
    &=& \cos(J\phi)\left[(i)^{J+m}+(i)^{J-m}\right]+\sin(J\phi)\left[(i)^{J-m}-(i)^{J+m}\right],\nonumber\\
    &&
\end{eqnarray}
Substituting Eq.~\eqref{B6} into Eq.~\eqref{B5},
\begin{eqnarray}\label{B7}
    && e^{i \frac{\pi}{2} \hat J_z^2} e^{i\frac{\pi}{2}\hat J_x} e^{-i \phi \hat J_z}\left(|J, J\rangle +|J, -J\rangle\right)\nonumber\\
    &=& \left[\cos(J\phi)\!-\!\sin(J\phi)\right]\prod_{l=1}^{2J}\left(\frac{1}{\sqrt{2}}|\downarrow\rangle_{l}+\frac{i}{\sqrt{2}}|\uparrow\rangle_{l}
    \right)\nonumber\\
    &+& \left[\cos(J\phi)\!+\!\sin(J\phi)\right]\prod_{l=1}^{2J}\left(\frac{i}{\sqrt{2}}|\downarrow\rangle_{l}+\frac{1}{\sqrt{2}}|\uparrow\rangle_{l}
    \right).\nonumber\\
    &&
\end{eqnarray}
So, we have
\begin{eqnarray}\label{B8}
    && e^{-i\frac{\pi}{2}\hat J_x} e^{i \frac{\pi}{2} \hat J_z^2} e^{i\frac{\pi}{2}\hat J_x} e^{-i \phi \hat J_z}\left(|J, J\rangle +|J, -J\rangle\right)\nonumber\\
    &=& \left[\cos(J\phi)\!-\!\sin(J\phi)\right]\prod_{l=1}^{2J} |\downarrow\rangle_{l}+ \left[\cos(J\phi)\!+\!\sin(J\phi)\right]\prod_{l=1}^{2J} |\uparrow\rangle_{l}\nonumber\\
    &=& \left[\cos(J\phi)\!-\!\sin(J\phi)\right]|J,-J\rangle + \left[\cos(J\phi)\!+\!\sin(J\phi)\right]|J,J\rangle.\nonumber\\
    &&
\end{eqnarray}
Next, when $-J<k<J$, we assume $k>0$ without loss of generality,
\begin{eqnarray}\label{B9}
    && e^{i\frac{\pi}{2}\hat J_x} e^{-i \phi \hat J_z}\left(|J, k\rangle +|J, -k\rangle\right)\nonumber\\
    &=& e^{i\frac{\pi}{4}\left(\sum_{l=1}^{2J}\hat \sigma_x^{(l)}\right)} \left(\prod_{l=1}^{J+k} e^{-i\frac{\phi}{2}}|\uparrow\rangle_{l}\prod_{l=J+k+1}^{2J}e^{i\frac{\phi}{2}}|\downarrow\rangle_{l}\right)\nonumber\\
    &+& e^{i\frac{\pi}{4}\left(\sum_{l=1}^{2J}\hat \sigma_x^{(l)}\right)} \left(\prod_{l=1}^{J-k} e^{-i\frac{\phi}{2}}|\uparrow\rangle_{l}\prod_{l=J-k+1}^{2J}e^{i\frac{\phi}{2}}|\downarrow\rangle_{l}\right)\nonumber\\
    &=& \sum_{m=-k}^{k} \frac{\sqrt{C_{2k}^{k-m}}}{2^k}\left[(i)^{k+m}e^{i\phi k}+(i)^{k-m}e^{-i\phi k}\right]|J,m\rangle. \nonumber \\
\end{eqnarray}
Then,
\begin{eqnarray}\label{B10}
    && e^{i \frac{\pi}{2} \hat J_z^2} e^{i\frac{\pi}{2}\hat J_x} e^{-i \phi \hat J_z}\left(|J, k\rangle +|J, -k\rangle\right)\nonumber\\
    &=& \sum_{m=-k}^{k} \frac{\sqrt{C_{2k}^{k-m}}}{2^k}(i)^{m^2} \left[(i)^{k+m}e^{i\phi k}+(i)^{k-m}e^{-i\phi k}\right]|J,m\rangle.\nonumber\\
    &&
\end{eqnarray}
Similar to Eq.~\eqref{B6}, we have
\begin{eqnarray}\label{B11}
    && (i)^{m^2} \left[(i)^{k+m}e^{i\phi k}+(i)^{k-m}e^{-i\phi k}\right]\nonumber\\
    &=& \cos(k\phi)\left[(i)^{k+m}+(i)^{k-m}\right]+\sin(k\phi)\left[(i)^{k-m}-(i)^{k+m}\right],\nonumber\\
    &&
\end{eqnarray}
Substituting Eq.~\eqref{B11} into Eq.~\eqref{B10},
\begin{eqnarray}\label{B12}
    && e^{i \frac{\pi}{2} \hat J_z^2} e^{i\frac{\pi}{2}\hat J_x} e^{-i \phi \hat J_z}\left(|J, k\rangle +|J, -k\rangle\right)\nonumber\\
    &=& \left[\cos(k\phi)\!-\!\sin(k\phi)\right]\prod_{l=1}^{2k}\left(\frac{1}{\sqrt{2}}|\downarrow\rangle_{l}+\frac{i}{\sqrt{2}}|\uparrow\rangle_{l}
    \right)\nonumber\\
    &&\prod_{l=2k+1}^{2J}
    \!\left(\frac{1}{\sqrt{2}}|\downarrow\rangle_{l}\!+\!\frac{1}{\sqrt{2}}|\uparrow\rangle_{l}\right)+\nonumber\\
    && \left[\cos(k\phi)\!+\!\sin(k\phi)\right]\prod_{l=1}^{2k}\left(\frac{i}{\sqrt{2}}|\downarrow\rangle_{l}+\frac{1}{\sqrt{2}}|\uparrow\rangle_{l}
    \right)\nonumber\\
    &&\prod_{l=2k+1}^{2J}
    \!\left(\frac{1}{\sqrt{2}}|\downarrow\rangle_{l}\!+\!\frac{1}{\sqrt{2}}|\uparrow\rangle_{l}\right).\nonumber\\
    &&
\end{eqnarray}
Finally, we obtain that
\begin{eqnarray}\label{B13}
    && e^{-i\frac{\pi}{2}\hat J_x} e^{i \frac{\pi}{2} \hat J_z^2} e^{i\frac{\pi}{2}\hat J_x} e^{-i \phi \hat J_z}\left(|J, k\rangle +|J, -k\rangle\right)\nonumber\\
    &=& \left[\cos(k\phi)\!-\!\sin(k\phi)\right](i)^{(J-k)^2}|J,-k\rangle \nonumber\\
    &+& \left[\cos(k\phi)\!+\!\sin(k\phi)\right](i)^{(J-k)^2}|J,k\rangle. \nonumber\\
    &&
\end{eqnarray}

Combining Eq.~\eqref{B8} and~\eqref{B13}, we can unify as Eq.~\eqref{Psi_nonDy} in the main text.

\end{document}